\documentclass[reprint,prl,superscriptaddress,showpacs,twocolumn]{revtex4}
\usepackage{units}
\usepackage{amsmath}
\usepackage{amssymb}
\usepackage{graphicx}
\usepackage{bm}
\usepackage{microtype,color}

\newcommand{\be}{\begin{equation}}
\newcommand{\ee}{\end{equation}}
\newcommand{\bea}{\begin{eqnarray}}
\newcommand{\eea}{\end{eqnarray}}

\begin{document}

\title{Control of energy density inside disordered medium by coupling to open or closed channels}

\author{Raktim Sarma}
\affiliation{Department of Applied Physics, Yale University, New Haven, CT, 06520, USA}
\author{Alexey G. Yamilov}
\email{yamilov@mst.edu}
\affiliation{\textls[-20]{Department of Physics, Missouri University of Science \& Technology, Rolla, Missouri 65409, USA}}
\author{Sasha Petrenko}
\affiliation{\textls[-20]{Department of Physics, Missouri University of Science \& Technology, Rolla, Missouri 65409, USA}}
\author{Yaron Bromberg}
\affiliation{Department of Applied Physics, Yale University, New Haven, CT, 06520, USA}
\author{Hui Cao}
\email{hui.cao@yale.edu}
\affiliation{Department of Applied Physics, Yale University, New Haven, CT, 06520, USA}

\date{\today}

\begin{abstract}

We demonstrate experimentally an efficient control of light intensity distribution inside a random scattering system. The adaptive wavefront shaping technique is applied to a silicon waveguide containing scattering nanostructures, and the on-chip coupling scheme enables access to all input spatial modes. By selectively coupling the incident light to open or closed channels of the disordered system, we not only vary the total energy stored inside the system by 7.4 times, but also change the energy density distribution from an exponential decay to a linear decay and to a profile peaked near the center. This work provides an on-chip platform for controlling light-matter interactions in turbid media.

\end{abstract}


\maketitle

It has long been known that in disordered media there are many fascinating counter-intuitive effects resulting from interferences of multiply scattered waves \cite{VanRossum_RMP,Akkermanbook}. One of them is the creation of transmission eigenchannels which can be broadly classified as open and closed \cite{Dorokhov1984,1986_Imry}. Existence of high-transmission (open) channels allows an optimally prepared coherent input beam transmitting through a lossless diffusive medium with order unity efficiency. Opposite to that, waves injected to low-transmission (closed) channels can barely penetrate the medium and are mostly reflected instead. In general, the penetration depth and energy density distribution of multiply scattered waves inside a disordered medium are determined by the spatial profiles of the transmission eigenchannels that are excited by the incident light. The distinct spatial profiles of open and closed channels suggest that selective coupling of incident light to these channels enables an effective control of total transmission and energy distribution inside the random medium \cite{Sarma15,Genack15}. Since the energy density determines the light-matter interactions inside a scattering system, manipulating its spatial distribution opens the door to tailoring optical excitations as well as linear and nonlinear optical processes such as absorption, emission, amplification, and frequency mixing inside turbid media. The potential applications range from photovoltaics \cite{SolarCell2,Wiersma_NatPho13}, white LEDs \cite{WhiteLED1} and random lasers \cite{Cheng14}, to biomedical sensing \cite{sensing} and radiation treatments \cite{ScientificReportsYoon15}.

In recent years there have been numerous theoretical and experimental studies on transmission eigenchannels \cite{MoskPRL,ChoiPRB,ChabanovNatComm13,Seng14,Genack15,MoskNatPhoReview}. While they can be deduced from the measured transmission matrix \cite{Popoff2010,Genack2012,YuPRL,Kim12}, it is difficult to directly probe their spatial profiles inside three-dimensional (3D) random media. So far, the open and closed channels are observed only with acoustic wave inside a two-dimensional (2D) disordered waveguide \cite{elasticwaves}, but controlling the energy density distribution has not been realized due to lack of an efficient wavefront modulator for acoustic wave or microwave. The advantage for optical wave is the availability of spatial light modulator (SLM) with many degrees of freedom, however, the commonly used samples in the optics experiment have an open slab geometry, making it impossible to control all input modes due to finite numerical aperture of the imaging optics. The incomplete control dramatically weakens the open channels \cite{Doug1}, although a notable enhancement of total transmission has been achieved \cite{Popoff2014,Kim12}. Furthermore, an enhancement of total energy stored inside a 3D scattering sample is reported \cite{Diffusionmode}, but direct probe and control of light intensity distribution inside the scattering medium are still missing.

In this Letter, we demonstrate experimentally control of energy density distribution inside a scattering medium.
Instead of the open slab geometry, we fabricate a silicon waveguide that contains scatterers and has reflecting sidewalls.
The intensity distribution inside the two-dimensional waveguide is probed from the third dimension. With a careful design of the on-chip coupling waveguide, we can access all the input modes. Such control of incident wavefront enables an order of magnitude enhancement of the total transmission or 50 times suppression.  A direct probe of light intensity distribution inside the disordered waveguide reveals that selective excitation of open channels results in an energy buildup deep inside the scattering medium, while the excitation of closed channels greatly reduces the penetration depth. Compared to the linear decay for random input fields, the optimized wavefront can produce an intensity profile that is either peaked near the center of the waveguide or decay exponentially with depth. The total energy stored inside the waveguide is increased 3.7 times or decreased 2 times.

The 2D waveguide structure was fabricated in a 220 nm silicon layer on top of 3 $\mu$m buried oxide by electron beam lithography and reactive ion etching \cite{Sarma15}. As shown in Fig. 1, air holes are randomly distributed within the waveguide whose sidewalls are made of photonic crystal to reflect light. At the probe wavelength $\lambda = 1.51$ $\mu$m, the transport mean free path $\ell = $ 2.5 $\mu$m is much less than the length $L$ = 50 $\mu$m  of the disordered waveguide, thus light transport is diffusive. The out-of-plane scattering, which enables a direct probe of light transport inside the random structure, can be treated as loss and the diffusive dissipation length is $\xi_a$ = 31 $\mu$m. The values of $\ell$ and $\xi_a$ were extracted from the measured intensity
distribution and intensity fluctuations inside the disordered waveguide for uncontrolled illumination \cite{Sarma1_15}. The waveguide of width $W$ = 15 $\mu$m supports $N$ = 56 transmission eigenchannels, among which $\sim$ 5 are open channels and the rest are closed channels. The total transmission for an uncontrolled illumination is about $4.8\%$.

The probe light is injected from the edge of the wafer to the waveguide. Due to the large mismatch of the refractive index between silicon and air, the light can be coupled only to the lower-order modes of the ridge waveguide. This limits the number of input modes that can be controlled by wavefront shaping. To increase the degree of input control, the coupling waveguide (lead) is tapered at an angle of $15^{\circ}$ [Fig. 1(a)]. The wider waveguide at the front end supports many more lower-order modes, which can be excited by the incident light and then converted to high-order modes by the tapering.

To select the initial width $W_1$ of the lead, we compute the degree of control for light field at the end of the tapered waveguide, that will be injected to the disordered waveguide. At the entrance of the lead, only low-order modes (up to $M_1-th$ order) of the waveguide (of width $W_1$) are excited with constant amplitude and random phase. We calculate the electric field at the end of the lead and construct the covariance matrix \cite{SI}. The sudden drop of the eigenvalues of the covariance matrix in Fig. 2(b) gives the number of significant eigenvalues, which corresponds to the number of independent spatial modes $M$ that are controlled by varying the input field [Fig. 2(b)]. We compute $M$ for different waveguide dimension and find $M = N$ as long as $M_1$ exceeds the number of transverse modes $N$ at the end of the lead (of width $W$). We then set $W_1$ = 330 $\mu$m for the fabricated sample in Fig. 1, and the number of waveguide modes that can be excited experimentally at the air/silicon interface is significantly larger than $N$. Thus all input modes to the disordered waveguide can be accessed by the incident fields to the lead with phase-only modulation.

\begin{figure}
\centering{\includegraphics[width=3.5in]{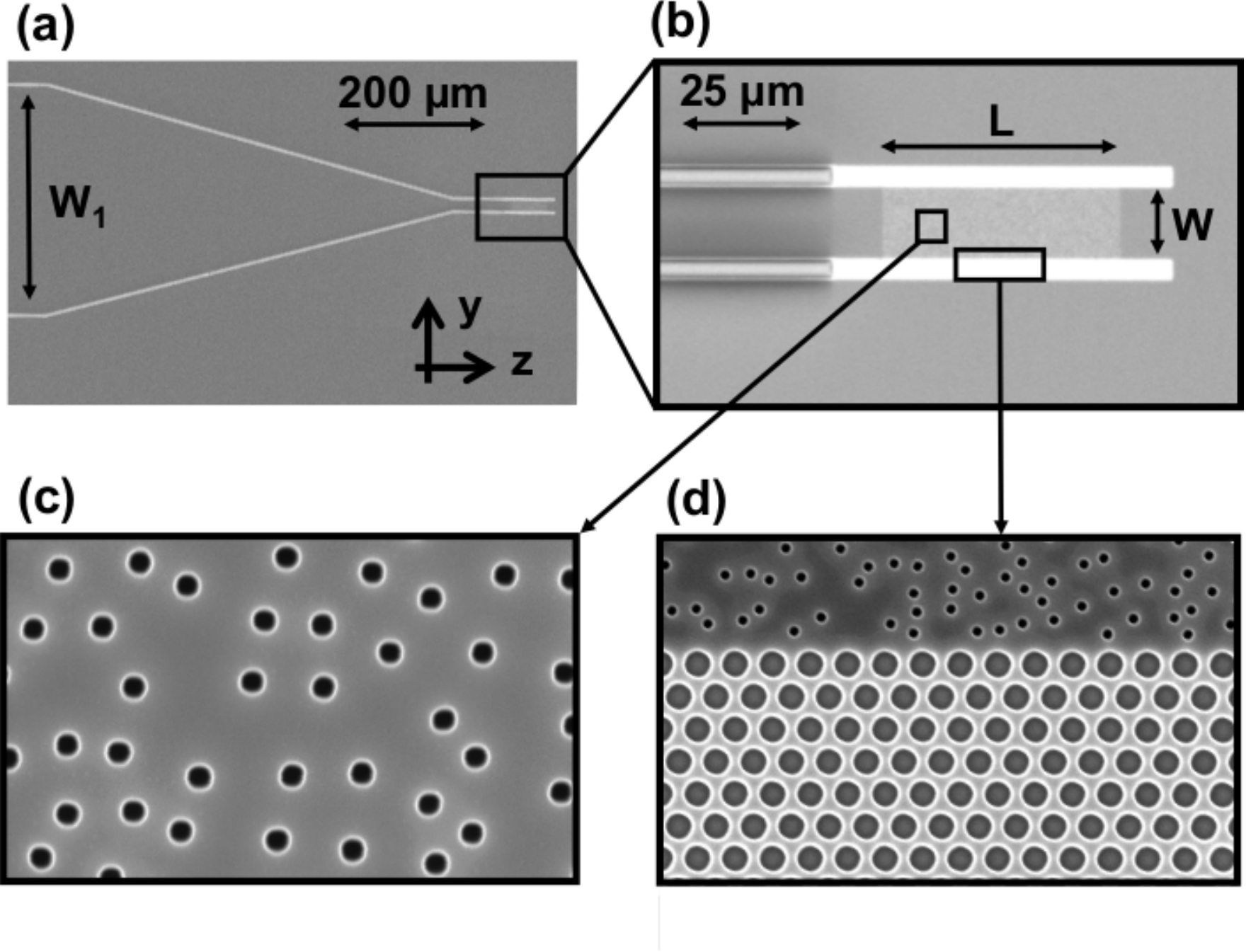}}
\caption{
\label{fig:sem}
{\bf On-chip disordered waveguide with a tapered lead.} (a) Top-view scanning electron micrograph (SEM) of a fabricated silicon waveguide. A ridge waveguide (lead) is tapered from the width $W_1$ = 330 $\mu$m at the edge of the wafer to the width $W$ = 15 $\mu$m, in order to increase the degree of control of the light that is injected to the disordered waveguide. (b) Magnified SEM of the disordered region of the waveguide that consists of a random array of air holes (diameter = 90 nm). (c) Magnified SEM showing the air holes distributed randomly within the waveguide with a filling fraction of 6 $\%$. (d) The sidewalls of the waveguide are made of a triangular lattice of air holes (diameter = 360 nm) with a lattice constant of 505 nm, which supports a full photonic bandgap at the wavelength $\lambda = 1.51$ $\mu$m.}
\end{figure}

The wavefront shaping experiment is shown schematically in Fig. 2(a) and detailed in the Supplementary Information \cite{SI}.
A monochromatic laser beam is phase modulated by a SLM, and then focused to the edge of the wafer by a microscope objective of numerical aperture (NA) 0.7. To produce a line of illumination at the input facet of the coupling waveguide, the SLM imposes phase modulation only in one direction, as shown by the 2D phase mask in Fig. 2(a). The light that is scattered out of plane by the random array of air holes is collected by an objective and projected to an InGaAs Camera to obtain the spatial distribution of the intensity, $I(y,z)$, inside the disordered structure [Fig 2(c)].

Two wavefront shaping approaches have been developed for transmission enhancement, one is based on the measurement of transmission matrix \cite{ChoiReview,ParkReview}, the other relies on the feedback \cite{VellekoopReview}. While the open channels can be obtained from the measured transmission matrix, the closed channels are subject to measurement noise due to nearly vanishing transmission.
Here we took the feedback approach, and adopted an optimization procedure based on the continuous sequential algorithm \cite{VellekoopReview} to control the energy density inside the disordered waveguide. The cost function $S$ is the ratio of light intensity integrated over an area in the front part of the waveguide to that in the back part [marked by two rectangles in Fig. 2(c)].

\begin{figure}
\centering{\includegraphics[width=3.5in]{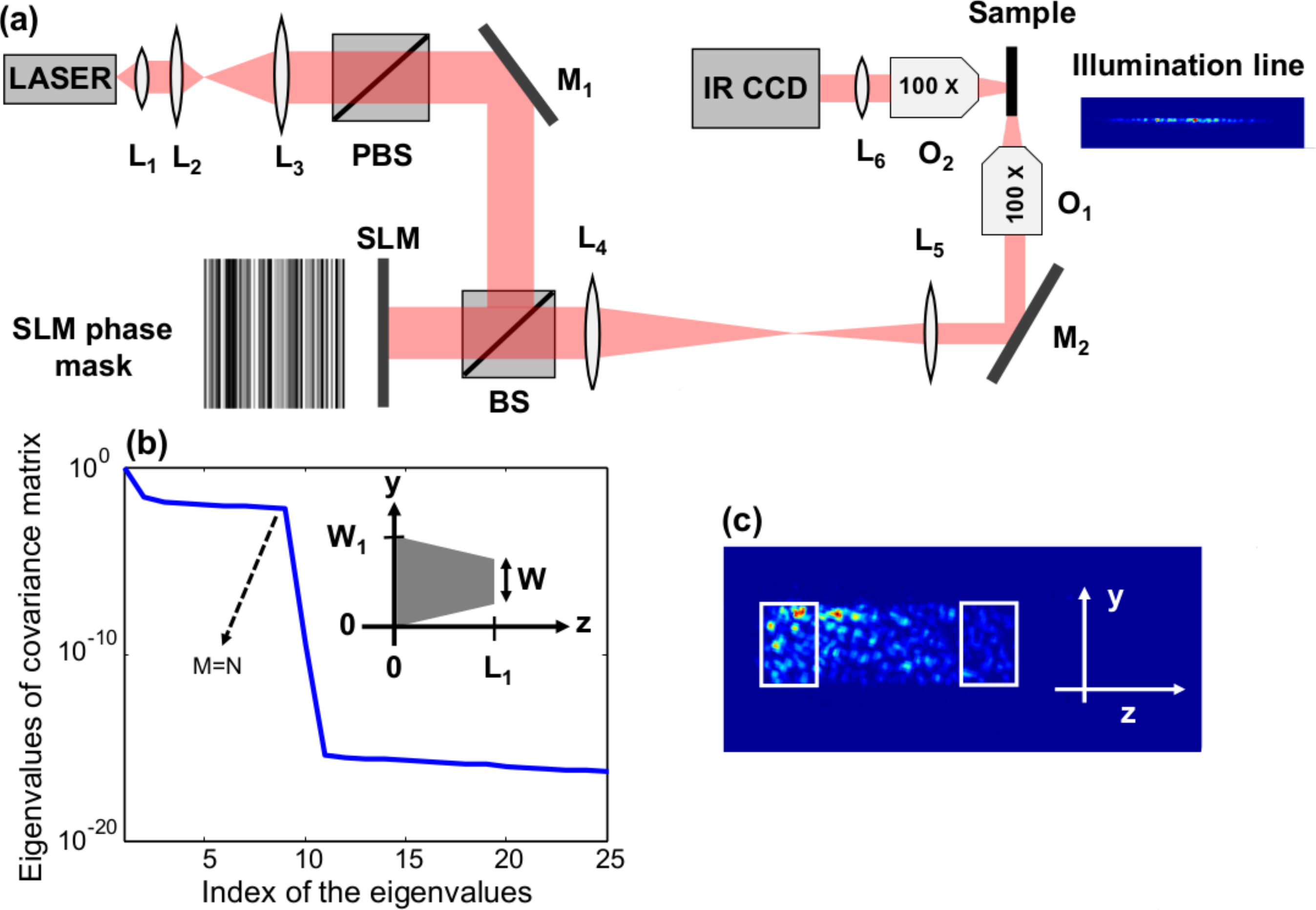}}
\caption{
\label{fig:experiment}
{\bf Wavefront shaping experiment to control intensity distribution inside a disordered waveguide.} (a) A schematic of the experimental setup. A laser (HP 8168F) output at $\lambda=1510$ nm is collimated (by lens $L_1$), expanded (by $L_2$, $L_3$) and linearly polarized (by a polarized beam splitter PBS) before being modulated by a phase-only SLM (Hamamatsu X10468). Two lens ($L_4$, $L_5$) are used to project the SLM plane to the pupil plane of an objective $O_1$ ($100 \times$, NA = 0.7), and the edge of the wafer is placed at the focal plane. The SLM imposes phase modulation only in one direction in order to generate a line at the front end of the coupling waveguide. A sample phase pattern and an image of the illumination line ($330 \times 1.1$ $\mu$m) are shown. The light scattered out of the sample plane is collected by another objective $O_2$ ($100 \times$, NA = 0.7) and imaged to an InGaAs camera (Xenics XEVA 1.7-320) by a tube lens ($L_6$). $M_1$ and $M_2$ are mirrors, BS is beam splitter.
(b) Semi-log plot of the eigenvalues of the covariance matrix $C(y,y')$ for the electric field $E_m(y,z=L_1)$ at the end of a coupling waveguide (lead). The inset is a schematic of the tapered lead.  $W_1 = 10$ $\mu$m, $W = 2.5$ $\mu$m, $L_1 = 20$ $\mu$m. The number of waveguide modes at $z=0$ is $N_1$ = 38, among them the low-order modes (from 1 to $M_1$ = 12) are excited by the incident field. The sudden drop of eigenvalues gives the number of significant eigenvalues, $M$ = 9, which gives the independent degrees of freedom at $z=L$. It is equal to the number of waveguide modes at $z = L_1$, $N$ = 9, since $M_1 > N$.
(c)  An image of the spatial distribution of light intensity inside the disordered waveguide for a random input wavefront. The spatial resolution is about 1.1 $\mu$m. The ratio $S$ of the integrated intensities over the two rectangles at the front and back side of the waveguide is used as feedback for optimizing the input wavefront.}
\end{figure}

First we maximize $S$ to enhance light penetration into the scattering structure. Figure 3(b) shows the final intensity distribution $I(y,z)$ for the optimized input. In Fig. 3(e) we plot the cross-section-averaged intensity $I(z) = \int_0^W I(y,z) dy$, further averaged over four wavelengths and three initial phase patterns. $I(z)$ is peaked near the center of the disordered waveguide in Fig. 3(e), which is dramatically different from the monotonic decay with random input fields in Fig. 3(d). The later profile agrees to the prediction of the diffusion theory, and the slight deviation from a linear decay is caused by the out-of-plane scattering loss. The optimized $I(z)$ resembles the spatial profile of open channels, indicating the optimized wavefront couples light to the high-transmission eigenchannels.

Next we minimize $S$ by adapting the input wavefront, and the resulting intensity distribution is presented in Fig. 3(c).
The cross-section-averaged intensity $I(z)$ in Fig. 3(f) exhibits a much faster decay with depth than the random input.
Moreover, the decay is clearly exponential, resembling the spatial profile of closed channels. Despite the presence of measurement noise, the optimized wavefront couples effectively  to the low-transmission eigenchannels.

To confirm the experimental results, we numerically simulate a 2D disordered waveguide with all parameters equal to the experimental values. The phase-only modulation is imposed to the input wavefront to optimize the same cost function $S$ with the continuous sequential algorithm (details in Supplementary Information)\cite{SI}. The solid curves in Fig. 3(d,e,f) represent the simulation results, which agree well to the experimental data.

\begin{figure*}
\centering{\includegraphics[width=\textwidth]{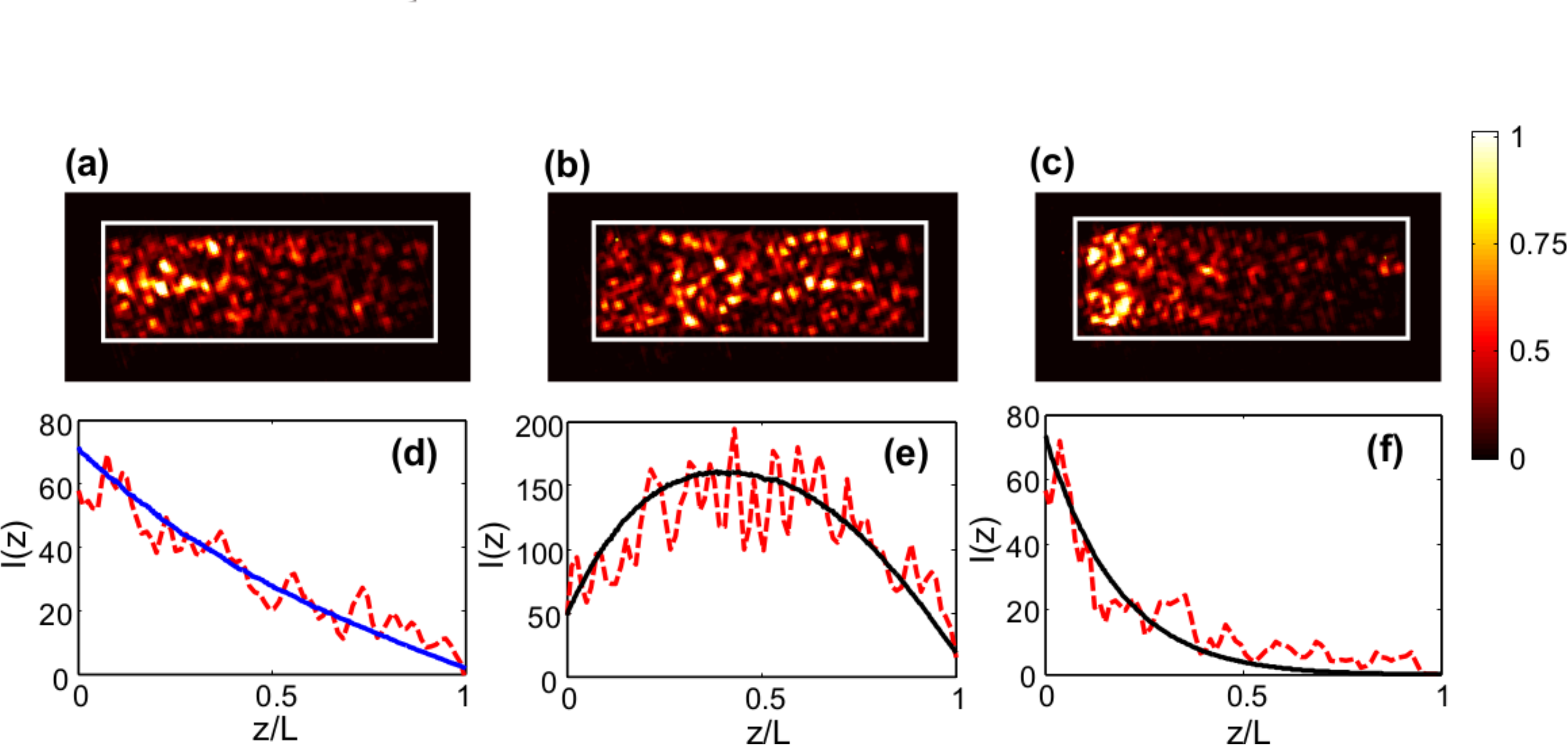}}
\caption{
\label{fig:Results}
{\bf Experimental control of intensity distribution inside the disordered waveguide.} (a, b, c) Two-dimensional intensity distribution $I(y,z)$ inside the disordered waveguide shown in Fig. 1 for (a) random input fields, (b) optimized input for maximum light penetration (maximizing $S$). (c) optimized input for minimum light penetration (minimizing $S$). (d, e, f) The cross-section-averaged intensity, $I(z)$, obtained from $I(y,z)$ in (a, b, c). Dashed lines are experimental data and solid lines are simulation results.}
\end{figure*}

By projecting the optimized fields to the transmission eigenchannels, we obtain the contributions from individual channels.
Figure 4(a) plots the weight $w$ of each channel as a function of the transmission eigenvalue $\tau$ in the case of maximizing the cost function $S$ [Fig. 3(b,e)]. In comparison to a random input field which has equal contributions from all channels $w(\tau) = 1/N$, the optimized field for maximum $S$ has greatly enhanced contributions from the high transmission channels and reduced contributions from the low-transmission channels [Fig. 4(a)]. While the maximum transmission channel has the largest weight, a few channels with slightly lower transmission also make significant contributions. Thus the energy density distribution $I(z)$ is slightly lower than that of the maximum transmission channel, and shifted a bit towards the front end of the waveguide [Fig. 4(b)]. As shown in Fig. 4(a), the weight $w(\tau)$ increases exponentially with $\tau$, in contrast to the linear increase of $w$ with $\tau$ in the case of focusing (maximizing intensity of a single speckle) through a random medium. This difference indicates maximizing $S$ is more efficient in enhancing the contribution of the maximum transmission channel over all other channels.

When $S$ is minimized [Fig. 3(c,f)], the weights of high-transmission channels are strongly suppressed, especially the highest transmission channel [Fig. 4(c)]. While many low-transmission channels have slightly increased weights as compared to the random input field, none of them becomes dominant. Consequently, the energy density distribution $I(z)$ decays exponentially, but the decay rate is slower than that of the minimum transmission channel [Fig. 4(d)].

The numerical simulation confirms that our wavefront shaping experiment results in selective coupling of the input light to open or closed channels, which leads to distinct intensity distribution inside the scattering waveguide. The total transmission is increased from $\sim 4.8\%$ (for random input fields) to $\sim 47\%$ (when $S$ is maximized), and the total energy inside the disordered structure is enhanced 3.7 times. The minimization of $S$ makes the total transmission drop to $\sim 0.1\%$, and the total energy inside is reduced by a factor of 2.

Finally we compare numerically the feedback-based approach to the transmission-matrix approach by computing the transmission eigenchannels from the field transmission matrix. With phase-only modulation, the input field for a transmission eigenchannel is decomposed by the waveguide modes, and the amplitude of the decomposition coefficients are set to a constant. The removal of amplitude modulation mixes the maximum transmission channel with other channels, as seen in Fig. 4(a). While the weight of the maximum transmission channel decreases from unity to $\pi/4$ \cite{Opt2008}, all other channels have a constant weight $(1-(\pi/4))/(N-1)$. The cross-section-averaged intensity distribution $I(z)$ is nearly identical to that obtained by maximizing $S$ [Fig. 4(b)]. Similarly, elimination of amplitude modulation from the minimum transmission channel introduces contributions from all other channels [Fig. 4(c)]. Their weights are equal (independent of their transmission), albeit smaller than that of the minimum transmission channel. Consequently, $I(z)$ displays a rapid decay at the shallow depth, due to the dominant contribution from the minimum transmission channel; it is followed by a much slower decay at large depth due to the contributions of the remaining channels including the highly transmitting ones. The total transmission is $\sim 1\%$, approximately an order of magnitude higher than that obtained by minimizing $S$. This is attributed to the stronger suppression of the higher transmission channels by the feedback approach, i.e., the higher the transmission eigenvalue, the lower the weight. Therefore, with phase-only modulation of incident wavefront, the feedback approach is far more efficient in minimizing the total transmission than the transmission-matrix approach.

\begin{figure}
\centering{\includegraphics[width=3.5in]{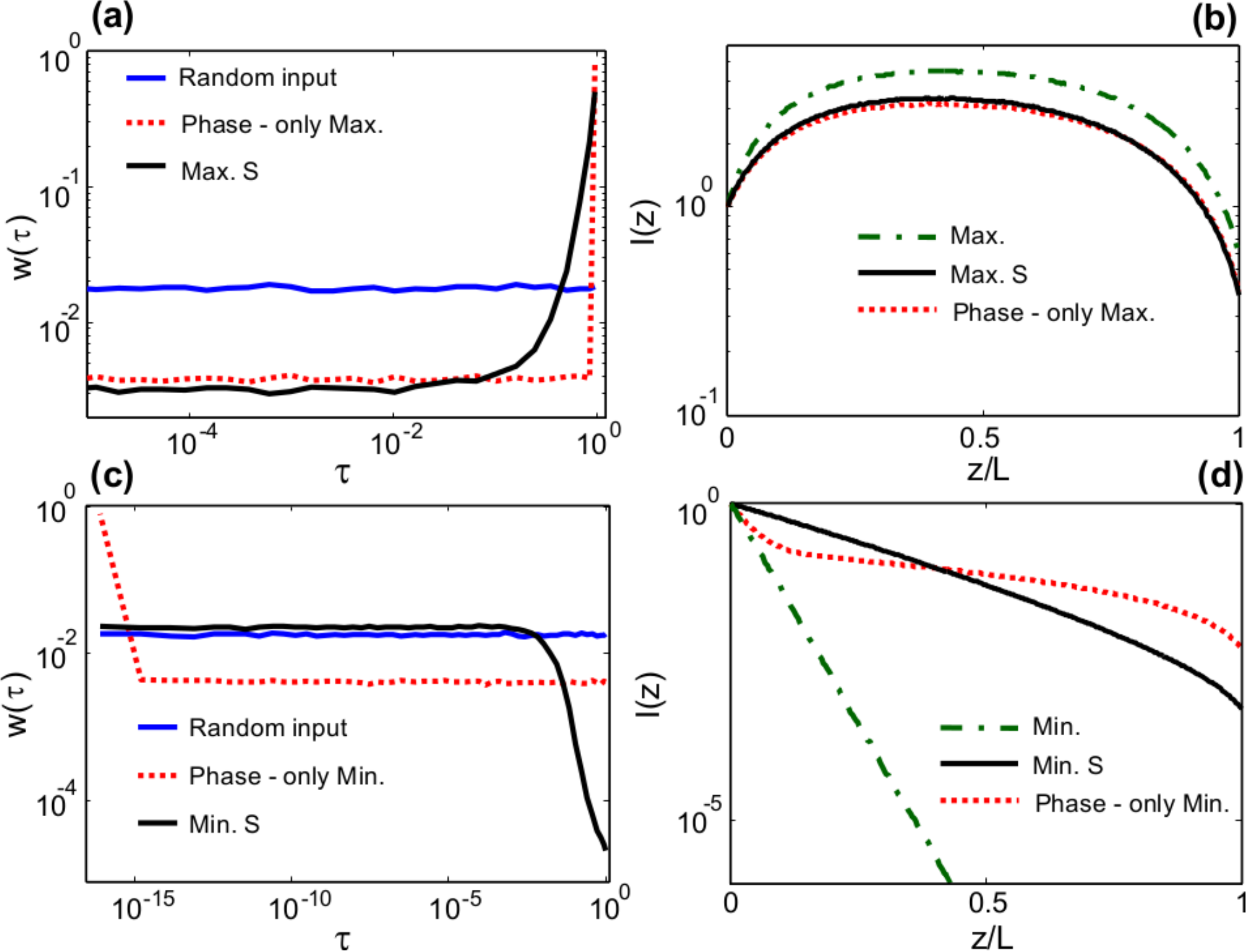}}
\caption{
\label{fig:simulations}
{\bf Numerical simulation of wavefront shaping experiment.} (a,c) Weight $w(\tau)$ of each transmission eigenchannel in the input field obtained by maximizing (a) or minimizing (c) light penetration into the disordered waveguide with the cost function $S$ (black solid line). For comparison, $w(\tau)$ for the random input field (blue solid line), and for the input field of the maximum (a) or minimum (c) transmission eigenchannel after removal of amplitude modulation (red dotted line) are also shown. (b,d) Cross-section-averaged intensity distribution $I(z)$ for the maximized (b) or minimized (d) $S$ (black solid line), as well as the maximum (b) or minimum (d) transmission channel with (green dash-dotted line) and without amplitude modulation (red dotted line).
}
\end{figure}

In summary, we apply the adaptive wavefront shaping technique to on-chip disordered nanostructures. A careful design of the coupling waveguide enables access to all input modes, thus allowing us to reach the maximum or minimum transmission that is achievable with phase-only modulation. Selective excitation of the open or closed channels results in the variation of light intensity distribution from an exponential decay to a linear decay and to a profile peaked near the center of the random system. The coherent control of multiple-scattering interference leads to diverse transport behaviors in violation of the universal diffusion, highlighting the feasibility of controlling light-matter interactions in turbid media.

\begin{acknowledgments}

We acknowledge Chia-Wei Hsu, Douglas Stone, Hasan Yilmaz, Seng Fatt Liew and Brandon Redding for useful discussions. This work was supported by National Science Foundation under grants nos. DMR-1205307 and DMR-1205223. Facilities use was supported by YINQE and NSF MRSEC DMR-1119826.

\end{acknowledgments}

 \end{document}